\documentclass[a4paper,11pt]{article}
\pdfoutput=1

\usepackage{jheppub}
\usepackage{amsmath}
\usepackage[T1]{fontenc}
\usepackage[normalem]{ulem}

\RequirePackage{lineno}
\usepackage{epstopdf}
\newcommand{\too}{\rightarrow}

\newcommand{\psipp}{\psi(3686)}
\newcommand{\chicJ}{\chi_{cJ}}
\newcommand{\jpsi}{J/\psi}
\newcommand{\ppb}{p\bar{p}}
\newcommand{\GG}{\gamma\gamma}
\newcommand{\Br}{\mathcal{B}}

\title{\boldmath Observation of $\chi_{cJ}(J=0,1,2)\to p\bar{p}\eta\eta$}

\collaborationImg{\includegraphics[width=.12\textwidth,origin=c,angle=90]{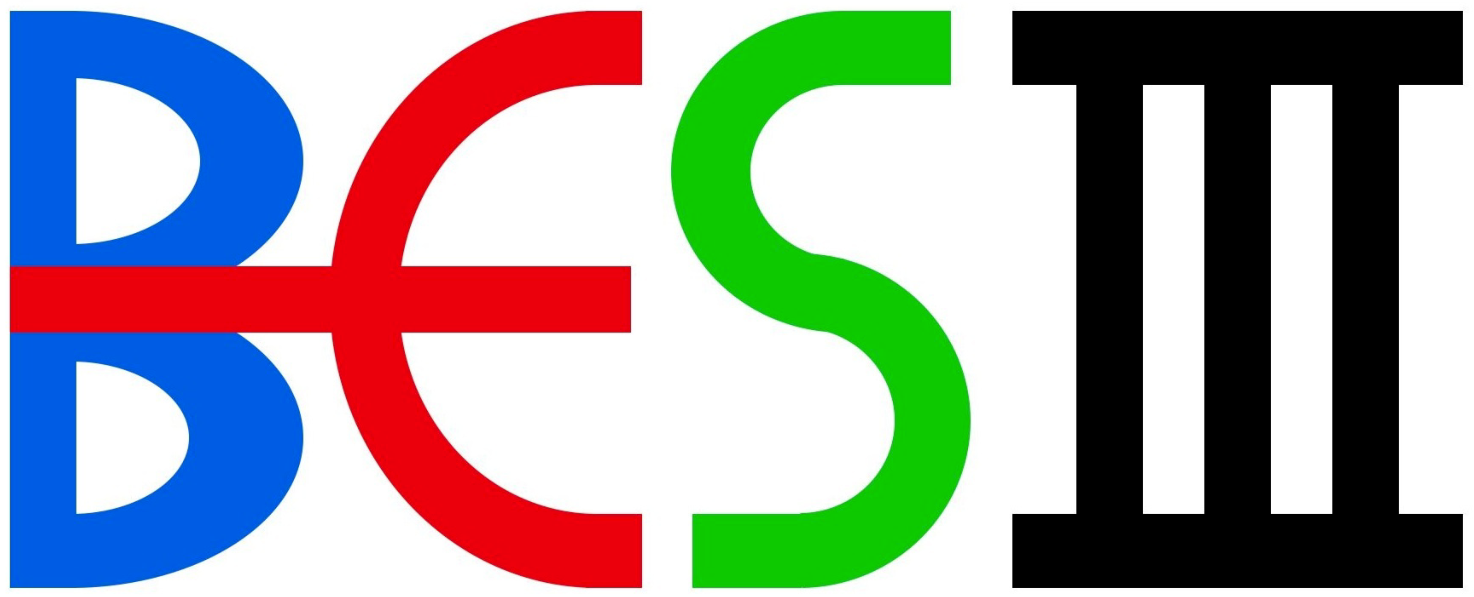}}

\collaboration{The BESIII collaboration}

\keywords{Charmonium, Branching fraction, BESIII}

\emailAdd{besiii-publications@ihep.ac.cn}

\arxivnumber{2505.12234}

\abstract{
Using $(2712.4\pm14.3)\times10^6$ $\psi(3686)$ candidates collected by the BESIII detector operating at the BEPCII storage ring, the decays  $\chi_{cJ}(J=0,1,2)\rightarrow p\bar{p}\eta\eta$ are observed for the first time through the radiative transition $\psi(3686)\to\gamma\chi_{cJ}$. The statistical significances for $\chi_{cJ}$ signals are all larger than 5$\sigma$. The branching fractions of $\chi_{c0,1,2}\to p\bar{p} \eta\eta$ are determined to be $({5.75 \pm 0.59 \pm 0.42}) \times 10^{-5}$, $({1.40 \pm 0.33 \pm 0.17}) \times 10^{-5}$, and $({2.64 \pm 0.40 \pm 0.27}) \times 10^{-5}$, respectively, where the first uncertainties are statistical and the second systematic. No evident resonant structures are found in the $p\bar{p}$ and $p\eta/\bar{p}\eta$ systems.}

\begin{document}
\maketitle
\flushbottom

%\linenumbers

\section{Introduction}

In the quark model, the $\chi_{cJ}(J=0,1,2)$ mesons are identified as $1^3P_J$ charmonium states. Because of parity conservation, they can only be produced in a two-photon exchange in direct $e^+e^-$ collisions, which are suppressed. The direct production $e^+e^-\to\chi_{c1}$ was observed by the BESIII Collaboration~\cite{ee_chic1}. However, the radiative decays of $\psipp$ into $\chicJ$ occur with significant branching fractions of approximately 10\% each~\cite{pdg2022}. Since 2008, BESIII has collected a large $\psipp$ data sample~\cite{psip_num_0912}, thereby providing an opportunity to further investigate the decays of $\chicJ$.

An intriguing enhancement near the $p\bar{p}$ threshold, referred to as the $X(p\bar{p})$, was discovered by the BES Collaboration in the radiative decay of $\jpsi\to\gamma p\bar{p}$~\cite{x1835} and subsequently confirmed by CLEO~\cite{cleo} and BESIII~\cite{B3}. However, no similar enhancement is found in the radiative decay $\Upsilon(1S)\to\gamma p\bar{p}$~\cite{1sppb} nor in the decay $\jpsi\to\omega p\bar{p}$~\cite{jjomg}. Many theories have been proposed to interpret the nature of this structure, including the quasibound nuclear baryonium~\cite{newb,Baryonium}, a multiquark resonance~\cite{hexaquark}, or an effect caused by final state interaction~\cite{fsi1,fsi2} near the $p\bar{p}$ production threshold. Therefore, further searches for an enhancement at this threshold in other hadronic final states are helpful to understand its properties.

Charmonium decays also provide an excellent venue to study some excited baryons, such as $N(1535)$ with $I(J^{P})=\frac{1}{2}({\frac{1}{2}}^-)$, which is observed with a mass close to the one predicted from the quark model~\cite{Vrana:1999nt}. However, its unexpectedly large branching fraction for $N(1535)\to p\eta$ remains puzzling~\cite{BESIII:2012ssm, BESIII:2013xkm}. BESIII has observed the $N(1535)$ via the decay of $\psi(3686) \to p\bar{p}\eta$~\cite{BESIII:2013xkm}. A search for $N(1535)$ via other charmonia, such as $\chicJ$ decays, can help understand the property of the $N(1535)$.

In this study, we present the first search for the decays $\chicJ\too\ppb\eta\eta$ $(J=0,1,2)$ via the radiative transition $\psi(3686)\to \gamma\chi_{cJ}$, based on an analysis of $(2712.4\pm14.3)\times10^6\ \psipp$ candidates~\cite{psip_num_0912}. Furthermore, the intermediate states in the $\ppb\eta\eta$ system are investigated.

\section{BESIII detector and Monte Carlo simulation}

The BESIII detector~\cite{Ablikim:2009aa} records $e^+ e^-$ collisions provided by the BEPCII storage ring~\cite{CXYu_bes3}. The cylindrical core of the BESIII detector covers 93\% of the full solid angle and consists of a helium-based multilayer drift chamber (MDC), a plastic scintillator time-of-flight system (TOF), and a CsI(Tl) electromagnetic calorimeter (EMC), which are all enclosed in a superconducting solenoidal magnet providing a 1.0~T magnetic field. The magnet is supported by an octagonal flux-return yoke with modules of resistive plate muon counters (MUC) interleaved with steel. The charged-particle momentum resolution at 1~GeV/$c$ is 0.5\%, and the  d$E/$d$x$ resolution is 6\% for electrons from Bhabha scattering at 1~GeV. The EMC measures photon energy with a resolution of 2.5\% (5\%) at 1~GeV in the barrel (end-cap) region. The time resolution of the TOF barrel part is 68~ps, while that of the end-cap part is 110~ps. The end-cap TOF system was upgraded in 2015 using multi-gap resistive plate chamber technology, providing a time resolution of 60~ps, which benefits 86$\%$ of data~\cite{tof_a,tof_b,tof_c}.

Monte Carlo (MC) simulated data samples are produced with a {\sc geant4}-based~\cite{geant4} software package, which includes the geometric description~\cite{detvis} of the BESIII detector and the detector response. These samples are used to optimize the event selection criteria, estimate the signal efficiency, and determine the level of background. The simulation models the beam energy spread and initial-state radiation in the $e^+e^-$ annihilation using the generator {\sc kkmc}~\cite{kkmc_a,kkmc_b}. The inclusive MC sample includes the production of the $\psi(3686)$ resonance, the initial-state radiation production of the $J/\psi$ meson, and the continuum processes incorporated in {\sc kkmc}. Particle decays are generated by {\sc evtgen}~\cite{evtgen_a,evtgen_b} for the known decay modes with branching fractions taken from the Particle Data Group  (PDG)~\cite{pdg2022} and {\sc lundcharm}~\cite{lundcharm_a,lundcharm_b} for the unknown ones. Final-state radiation from charged final-state particles is included using the {\sc photos} package~\cite{photos}.

The signal MC samples are generated by using the phase space (PHSP) and P2GC0/1/2 models. The P2GC0/1/2 generator models are set according to the quantum number and angular distribution of $\psipp \to \gamma \chicJ(J=0,1,2)$, respectively~\cite{chicjgen1,chicjgen2}. The other processes are generated by the PHSP model.

\section{Event selection}

In this analysis, the $\eta$ candidate is reconstructed from a $\GG$ final state. Therefore, the final state of interest is $\ppb5\gamma$. Hence, at least one positively charged and one negatively charged track are required. Charged tracks are required to originate within 10 cm from the interaction point in the $z$ direction and less than 1 cm in the plane perpendicular to the beam, and be within a range of $\left|\cos\theta\right|<0.93$, where $\theta$ is the polar angle with respect to the MDC symmetry axis. The combined information of the $\mathrm{d}E/\mathrm{d}x$ and TOF is used to calculate particle identification (PID) probabilities for the pion, kaon, and proton hypothesis, respectively, and the particle type with the highest probability is assigned to the corresponding track. Finally, the events with exactly two charged tracks, one proton and one anti-proton, are retained.

Photon candidates are selected using showers in the EMC. The deposited energy of each shower must be greater than 25 MeV in the barrel region ($|\cos\theta|<0.80$) or greater than 50\,\text{MeV} in the end-cap region ($0.86<|\cos\theta|<0.92$). To suppress electronic noise and energy deposition not associated to the event, the EMC cluster timing from the reconstructed event start time is further required to satisfy $0\leq t\leq$ 700 ns. The number of photon candidates is required to be at least five.

In order to suppress the remaining background and to improve the mass resolution, a four-constraint (4C) kinematic fit is performed with the $\psipp\to5\gamma\ppb$ hypothesis by constraining the total four-momentum of the final state particles to that of the colliding beams. If there is more than one combination surviving the 4C kinematic fit, the one with the least $\chi^{2}$ is chosen. Furthermore, the $\chi^2$ of the 4C kinematic fit is required to be less than 35, which is obtained by optimizing the figure-of-merit $\mathrm{FOM}=S/\sqrt {S+B}$ ~\cite{punzi}, where $S$ is the number of MC signal events normalized to the data from $\chi_{c0}\to \ppb\eta\eta$ and $B$ is obtained from the normalized inclusive MC sample. Moreover, in order to suppress the background from the final states with a non-nominal photon number, we perform a 4C kinematic fit by looping over all the four, five, and six-photon combinations, respectively. Four-photons final state are a background when they are combined with an additional fake photon, which can easily be produced by the EMC of BESIII. The combinations with the least $\chi^2$ are chosen for different photon number hypotheses. Then, we require $\chi^{2}(5\gamma p\bar{p})<\chi^{2}(4\gamma p\bar{p})$ and $\chi^{2}(5\gamma p\bar{p})<\chi^{2}(6\gamma p\bar{p})$. To suppress background with a final state that contains a $\pi^0$, such as $\chi_{cJ}\to p\bar{p}\pi^0\pi^0$, the invariant mass for all $\GG$ combinations is required to be out of the $\pi^0$ mass window, i.e., $\left|M(\gamma\gamma) -m_{\pi^0}\right|>15$ MeV$/c^2$, where $m_{\pi^0}$ is the $\pi^0$ nominal mass~\cite{pdg2022} and 15 MeV/$c^2$ is about three standard deviations of the resolution for the $\pi^0$.

\begin{figure}[htbp]		
\begin{center}
\includegraphics[width=0.5\textwidth]{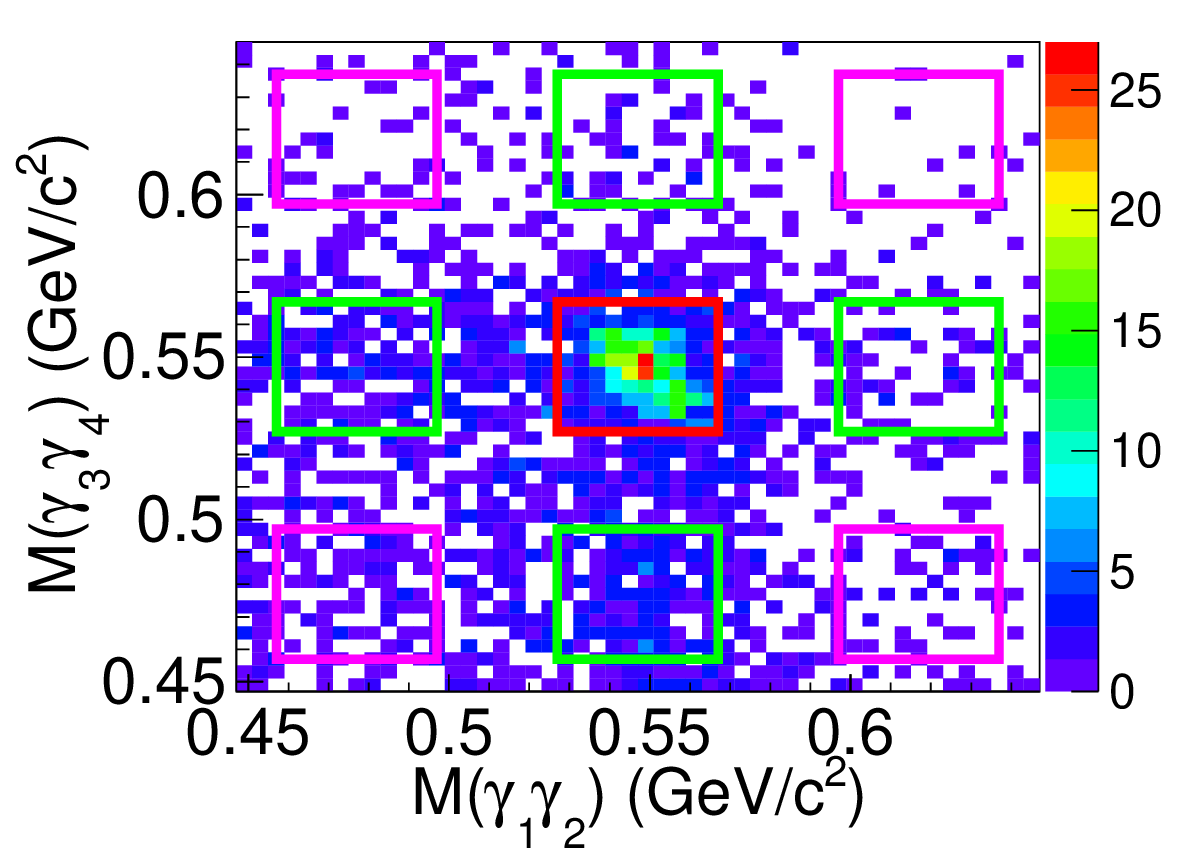}
\caption{The scatter plot of $M(\gamma_{1}\gamma_{2})$ versus $M(\gamma_{3}\gamma_{4})$ for data. The red box marks the $\eta\eta$ signal region, the green boxes mark the single-$\eta$ sideband regions and the purple boxes mark the non-$\eta$ sideband regions.}
\label{fig:2D_LLB}
\end{center}
\end{figure}

After applying all the above selection criteria, the $\eta\eta$ pair is selected from all four-photon combinations by minimizing
$\Delta =\sqrt{(M(\gamma_{1}\gamma_{2})-m_{\eta})^2+(M(\gamma_{3}\gamma_{4})-m_{\eta})^2}$, where $m_{\eta}$ is the $\eta$ nominal mass, and the subscripts are assigned to discriminate the different photon candidates. Figure~\ref{fig:2D_LLB} shows the two dimensional (2D) distribution of the invariant masses of the photon pairs. A clear $\eta\eta$ signal is seen in the central region. The signal region (the red box) is defined as $|M(\gamma_{1}\gamma_{2})-m_{\eta}|<20~\mathrm{MeV}/c^2$ and $|M(\gamma_{3}\gamma_{4})-m_{\eta}|<20~\mathrm{MeV}/c^2$. The single-$\eta$ sideband (SD1) regions are offset by $\pm70~\mathrm{MeV}/c^2$ along each axis (4 green squares) and the non-$\eta$ sideband (SD2) regions are shifted by $\pm70~\mathrm{MeV}/c^2$ along both axis simultaneously (4 purple squares).

\begin{figure*}[htbp]
\includegraphics[width=1.04\textwidth]{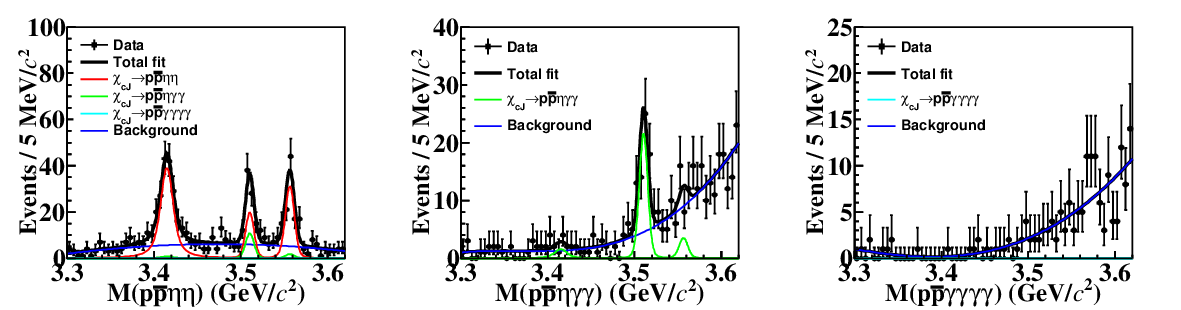}
\caption{Simultaneous fit to the $M(p\bar{p}\eta\eta)$ (left), $M(p\bar{p}\eta\gamma\gamma)$ (middle) and $M(p\bar{p}\gamma\gamma\gamma\gamma)$ (right) distributions of accepted candidate events in the $\chi_{cJ}$ mass region for $\eta\eta$ signal region (left), SD1 region (middle) and SD2 region (right).}
\label{fig:simultaneousFit}
\end{figure*}

Figure~\ref{fig:simultaneousFit} shows the $M(p\bar{p}\eta\eta)$, $M(p\bar{p}\eta\gamma\gamma)$ and $M(p\bar{p}\gamma\gamma\gamma\gamma)$ distributions of the survival event candidates in the $\eta\eta$, SD1 and SD2 regions. Significant $\chi_{cJ}$ signals are seen for the events in the $\eta\eta$ signal region, while there is also $\chi_{cJ}$ peaking background for the sideband events from SD1 region, but there is no significant $\chi_{cJ}$ peaking background for the sideband events from SD2 region. The potential remaining background from $\psi(3686)$ decays is investigated with the $\psi(3686)$ inclusive MC samples, using the event-type analysis tool TopoAna~\cite{zhouxy_topoAna}. It is found that although there are some surviving backgrounds from other $\psi(3686)$ decays, the distribution is smooth without significant $\chi_{cJ}$ peaks. To investigate the continuum background from direct $e^+e^-$ collisions without through $\psi(3686)$ decays, the same selection criteria are applied to the data samples collected at the center-of-mass energy of 3.650 GeV, 3.682 GeV, and 3.773 GeV. The total integrated luminosity of these samples amounts to 3931 $\mathrm{pb^{-1}}$, while the $\psipp$ data sample corresponds to an integrated luminosity of 3877 $\mathrm{pb^{-1}}$. Only a few events are found, which is less than one percent of the selected candidates, so the continuum background is negligible. Therefore, the contribution from the $\chi_{cJ}$ peaking background can be estimated using normalized sideband events.

\section{\label{Sec:BR_determined}Signal yields}
The signal yields for $\chi_{c0,1,2}$ are obtained by performing a simultaneous fit to the mass spectrums of $M(p\bar{p}\eta\eta)$, $M(p\bar{p}\eta\gamma\gamma)$, $M(p\bar{p}\gamma\gamma\gamma\gamma)$ for the $\eta\eta$ signal and both sideband regions (SD1 and SD2). The signal shape of the $\chi_{cJ}$ is described by Breit-Wigner (BW) functions $\text{BW}(m,\Gamma)$ convolved with a Gaussian function accounting for the detector resolution whose parameters are left as free and shared by all $\chi_{cJ}$ signals. The mass ($m$) and width ($\Gamma$) in the BW function are fixed to those in the PDG~\cite{pdg2022}. The smooth background is described with a $2^{\rm nd}$-order Chebychev polynomial function due to its smooth shape. The $\chicJ$ signals from SD1 and SD2 are also described by BW functions $\text{BW}(m,\Gamma)$ convolved with a Gaussian function, whose parameters are the same as those used above. The number of $\chi_{cJ}$ peaking background events in the signal region is estimated to be $\frac{1}{2}$$N$(SD1)-$\frac{1}{4}$$N$(SD2) by assuming the single-$\eta$ and non-$\eta$ peaking background to be a linear distribution, where $N$(SD1) and $N$(SD2) are the number of $\chi_{cJ}$ events in the SD1 and SD2 regions, respectively. The $\chi_{cJ}\to \ppb\eta\eta$ signal yield $N_{\chi_{cJ}}^{\mathrm{obs}}$, $N$(SD1), $N$(SD2) for each of the $\chi_{cJ}$ and the smooth background in each region are floated in the fit. The fit results are shown in Fig.~\ref{fig:simultaneousFit} and summarized in Table~\ref{list_summary}.

\begin{table*}[htbp]
\begin{center}
\caption{ The $\chi_{cJ}\to \ppb\eta\eta$ signal yield $N_{\chi_{cJ}}^{\mathrm{obs}}$, $N$(SD1), $N$(SD2), significance (S), efficiency ($\epsilon$), and branching fractions $\Br(\chi_{cJ} \rightarrow \ppb\eta\eta)$. The first uncertainties are statistical and the second systematic.}
\label{list_summary}
\setlength{\tabcolsep}{8pt}
\begin{tabular}{l c c c c c c}
\hline
\hline
 & $N_{\chi_{cJ}}^{\mathrm{obs}}$ & $N$(SD1) & $N$(SD2) & S($\sigma$) & $\epsilon$(\%) & $\Br(\chi_{cJ} \to p\bar{p}\eta\eta)$($\times 10^{-5}$) \\
\hline
$\chi_{c0}$ & $180.4\pm18.4$ & $7.2 \pm 5.6$ & $0.0 \pm 1.4$ & 13.4 & 7.64 & $5.75\pm0.59\pm0.42$ \\
$\chi_{c1}$ & $50.9\pm11.9$ & $55.4 \pm 9.6 $ & $0.0 \pm 4.0$  & 5.4 &  8.88 & $1.40\pm0.33\pm0.17$  \\
$\chi_{c2}$ & $87.3\pm13.3$  &  $9.8  \pm 7.8 $ & $0.0 \pm 14.3$ & 9.6 &  8.42 & $2.64\pm0.40\pm0.27$ \\
\hline
\hline
\end{tabular}

\end{center}
\end{table*}

The statistical significances of $\chi_{c0}$, $\chi_{c1}$, and $\chi_{c2}$ decays are determined to be 13.4$\sigma$, 5.4$\sigma$, and 9.6$\sigma$, respectively, by comparing the difference of the likelihood with and without the respective signal component in the fit. The effect from the systematic uncertainty has been considered in determining the signal significance.

\begin{figure*}[htbp]
\begin{center}
\begin{minipage}[t]{0.32\linewidth}
\includegraphics[width=1\textwidth]{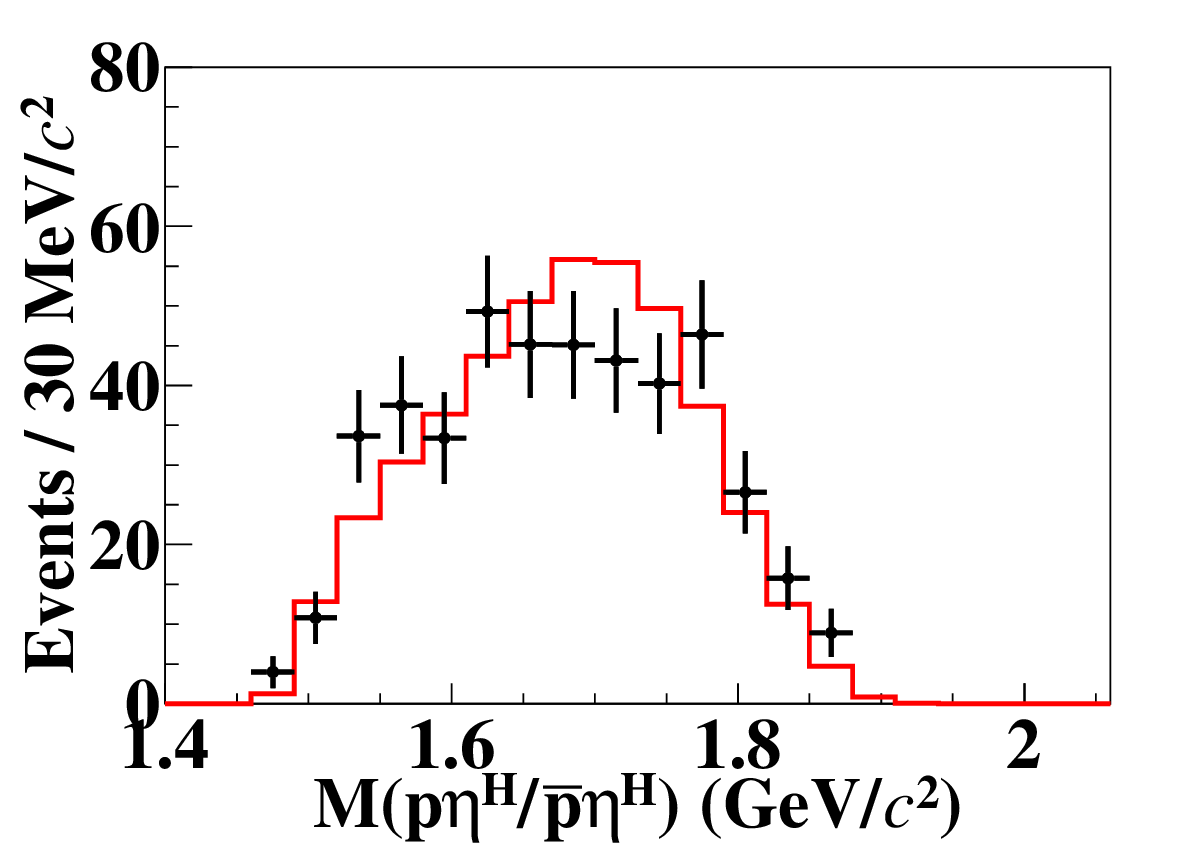}
\end{minipage}
\begin{minipage}[t]{0.32\linewidth}
\includegraphics[width=1\textwidth]{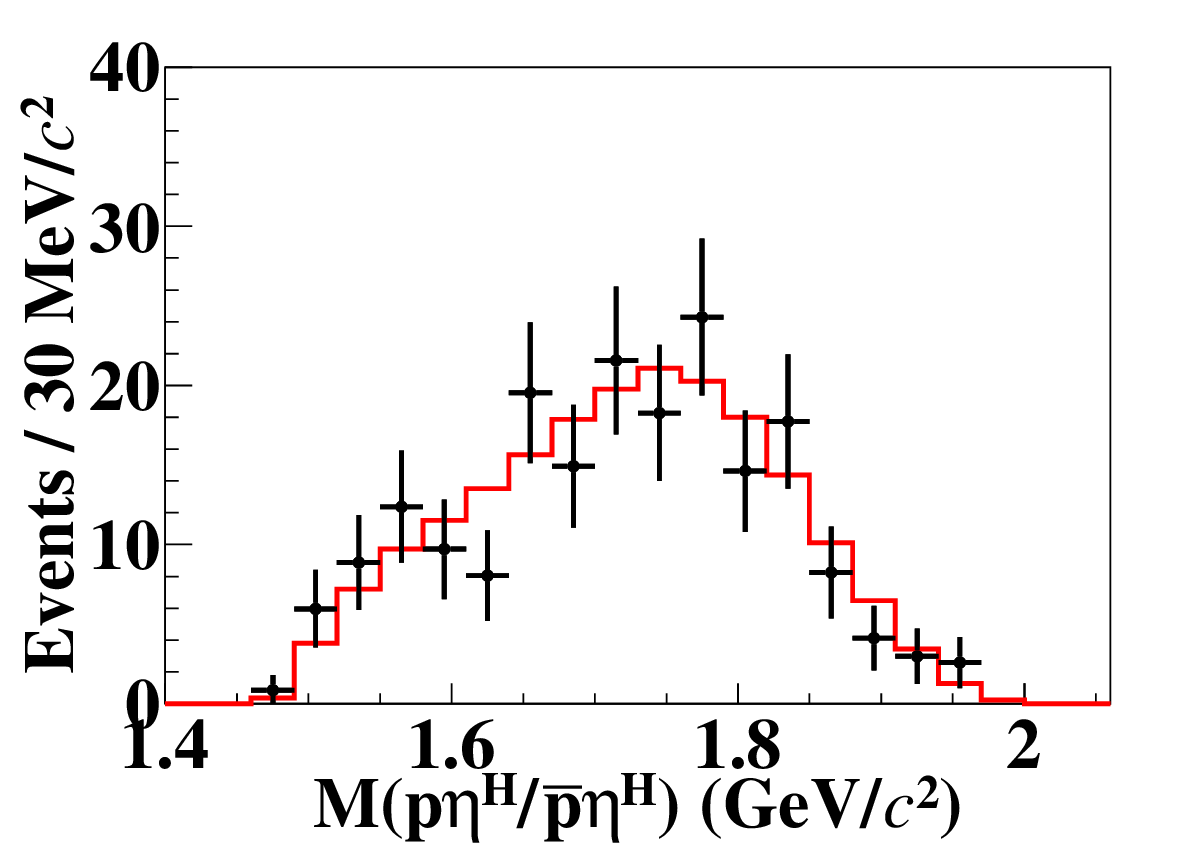}
\end{minipage}
\begin{minipage}[t]{0.32\linewidth}
\includegraphics[width=1\textwidth]{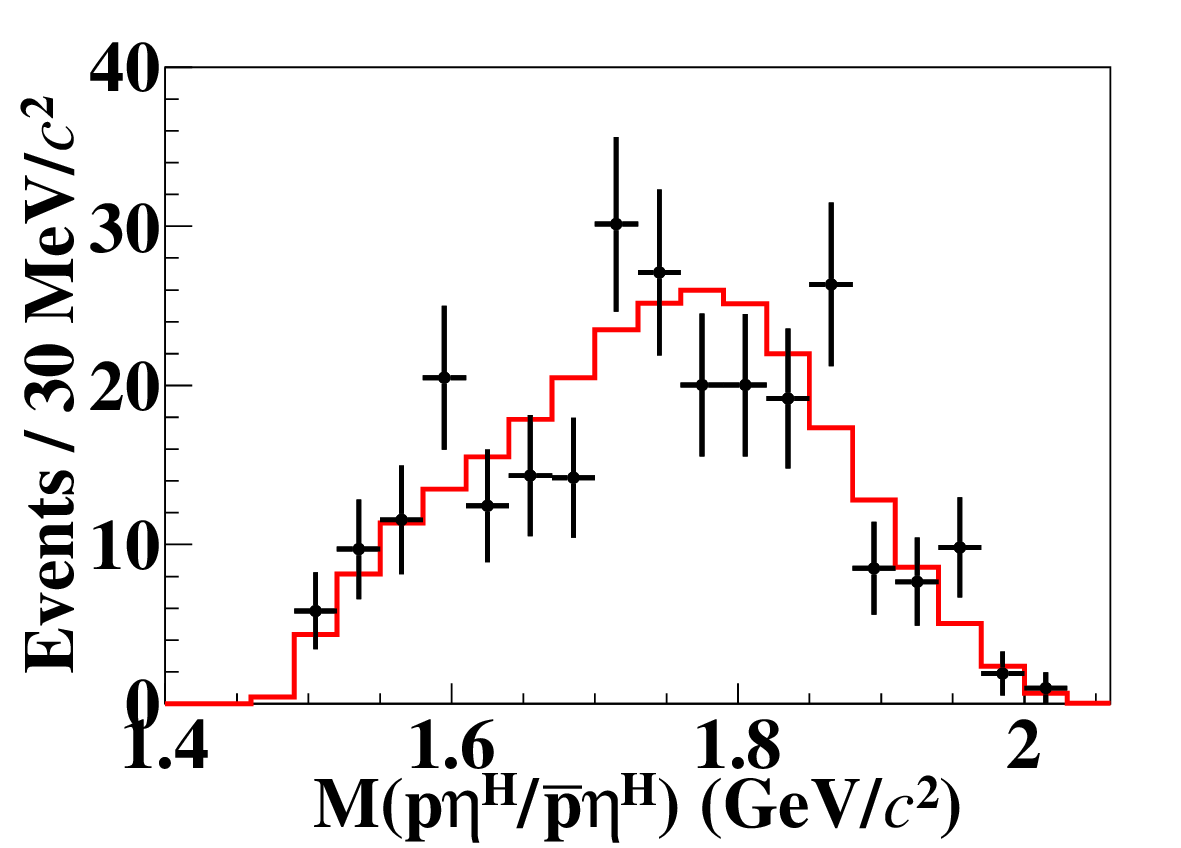}
\end{minipage}

\begin{minipage}[t]{0.32\linewidth}
\includegraphics[width=1\textwidth]{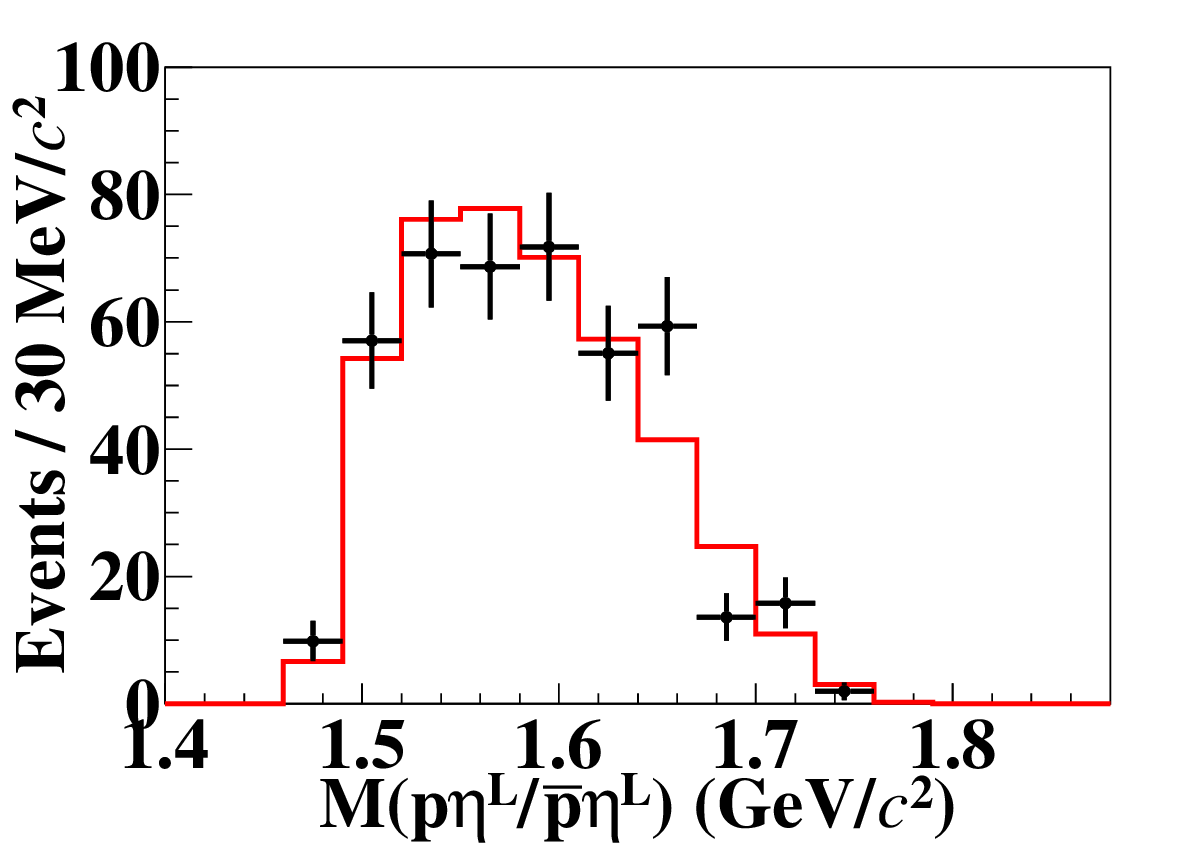}
\end{minipage}
\begin{minipage}[t]{0.32\linewidth}
\includegraphics[width=1\textwidth]{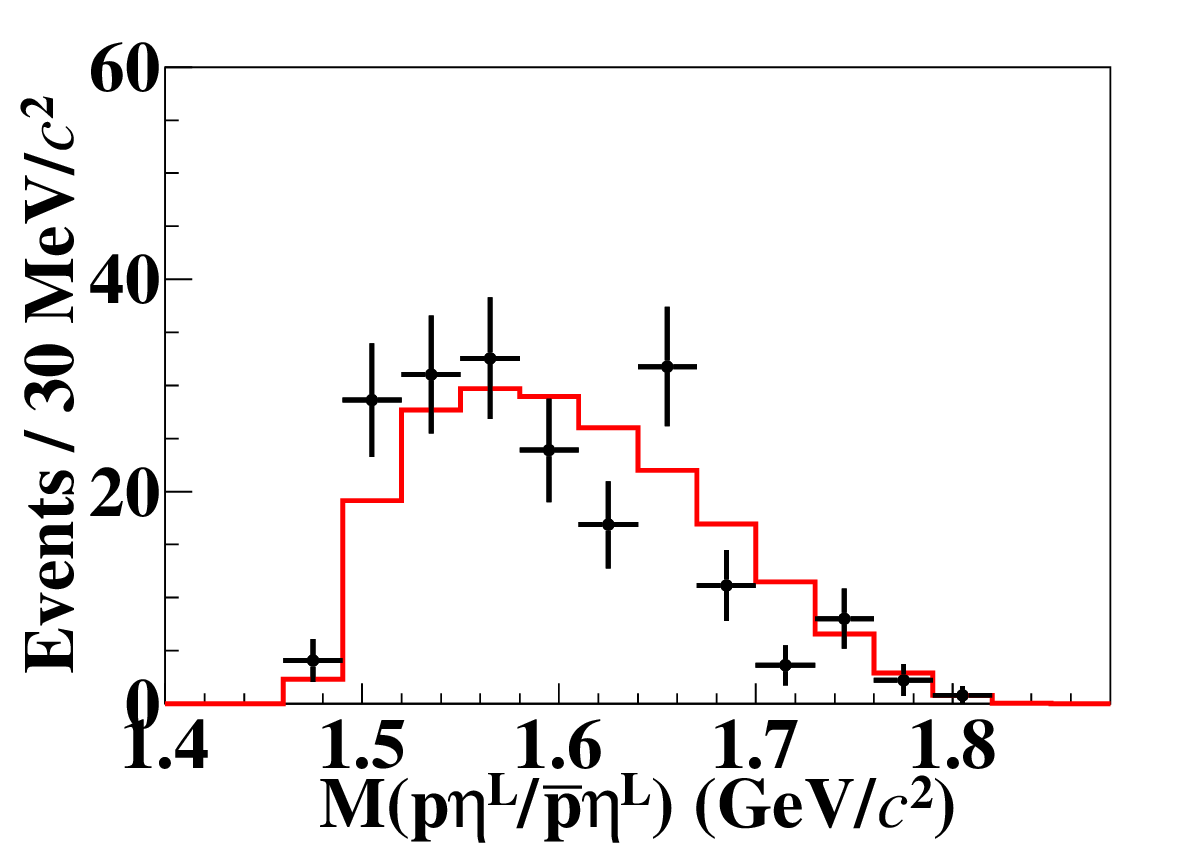}
\end{minipage}
\begin{minipage}[t]{0.32\linewidth}
\includegraphics[width=1\textwidth]{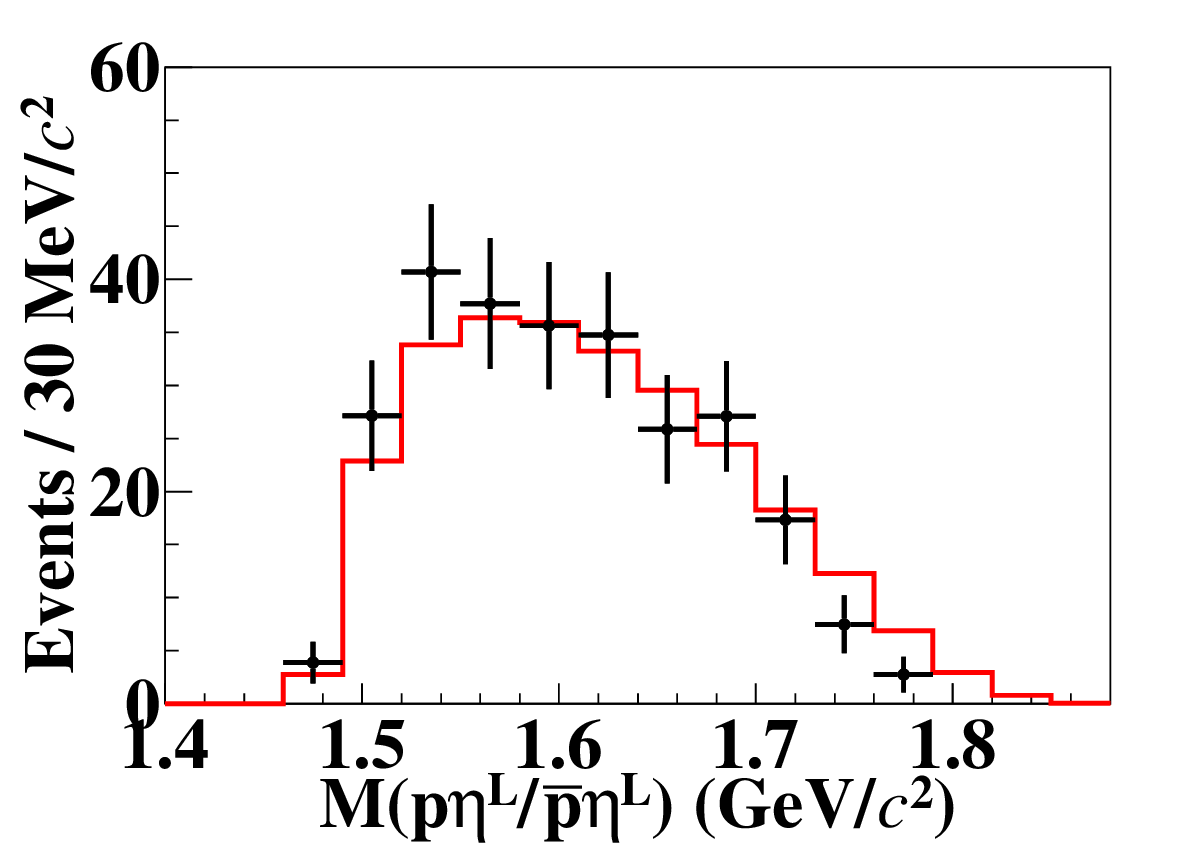}
\end{minipage}

\begin{minipage}[t]{0.32\linewidth}
\includegraphics[width=1\textwidth]{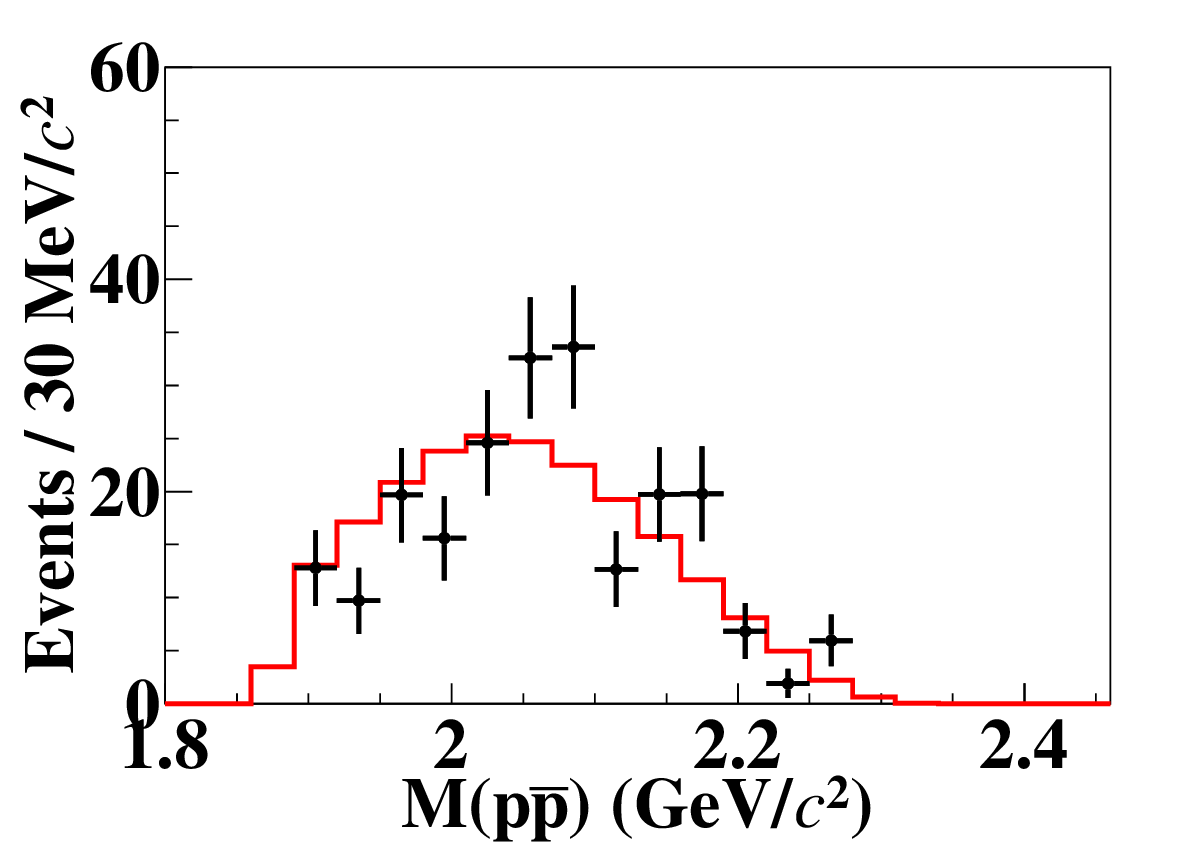}
\end{minipage}
\begin{minipage}[t]{0.32\linewidth}
\includegraphics[width=1\textwidth]{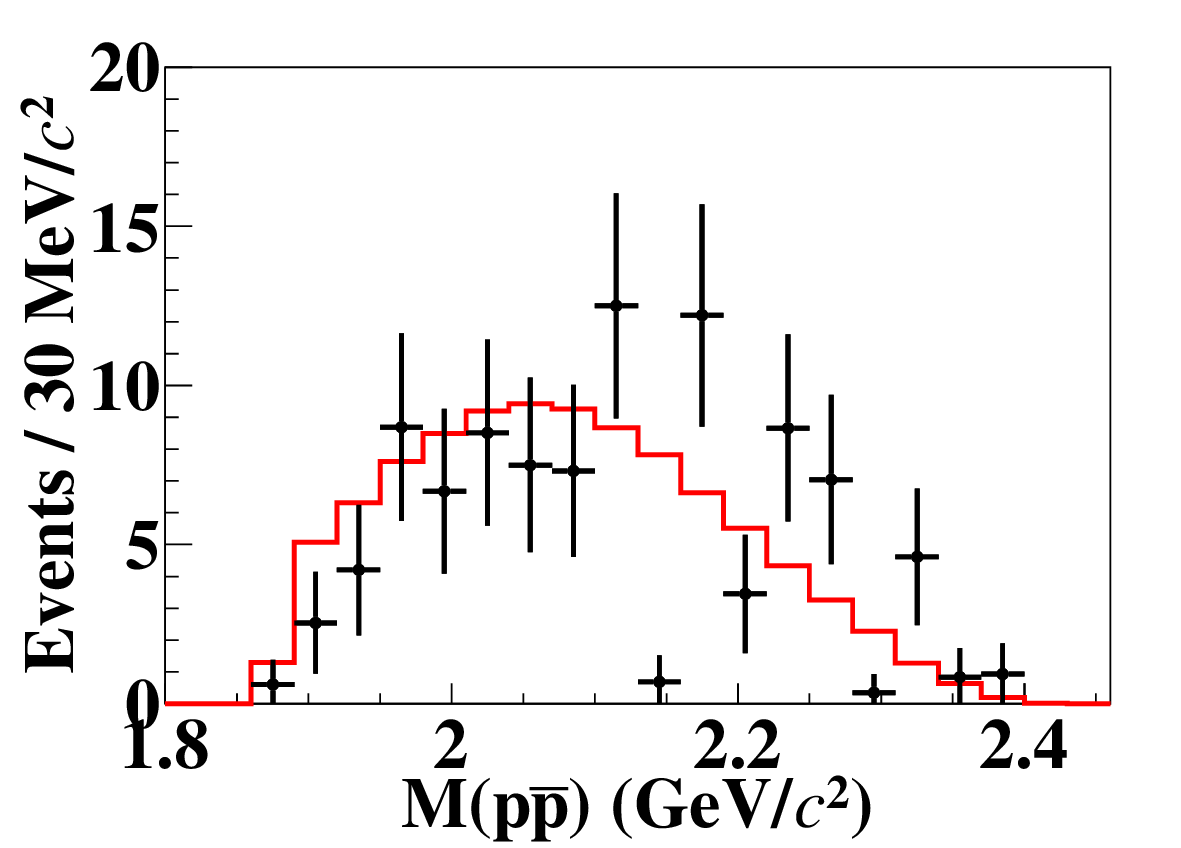}
\end{minipage}
\begin{minipage}[t]{0.32\linewidth}
\includegraphics[width=1\textwidth]{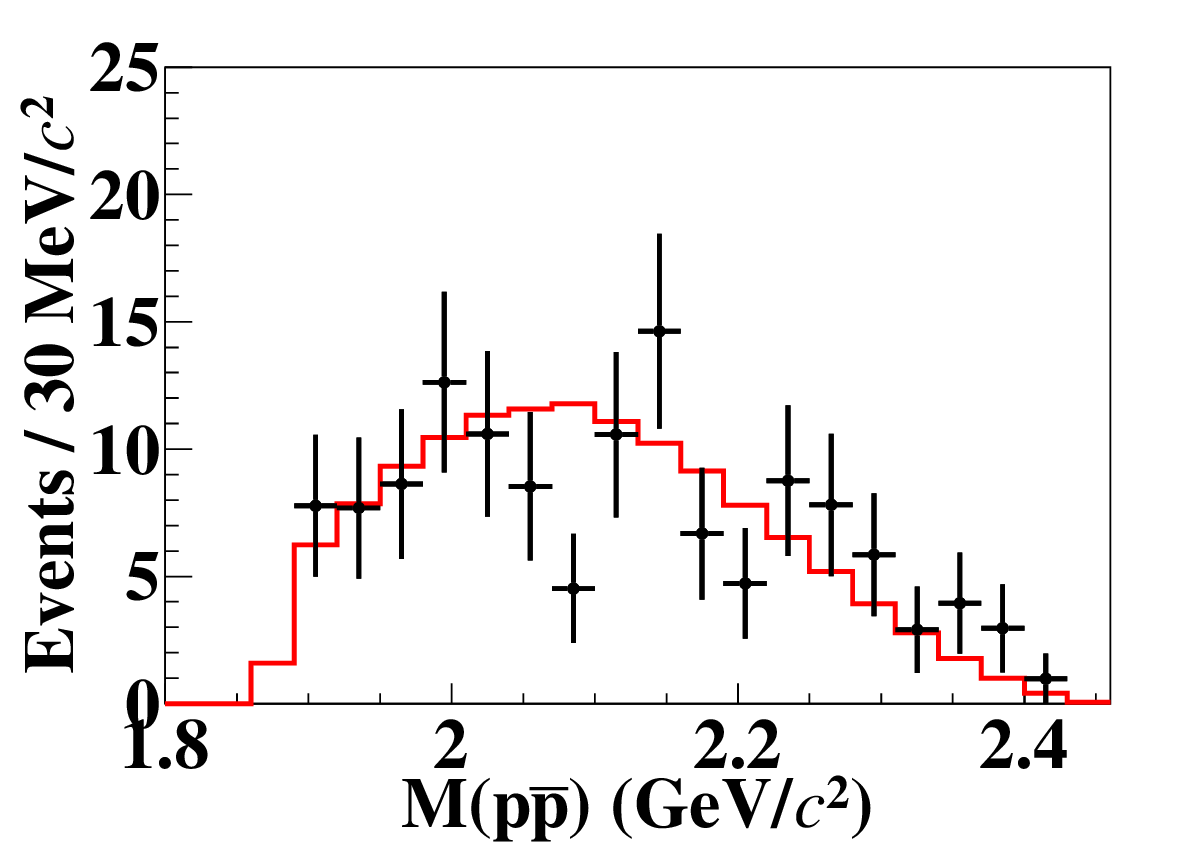}
\end{minipage}

\begin{minipage}[t]{0.32\linewidth}
\includegraphics[width=1\textwidth]{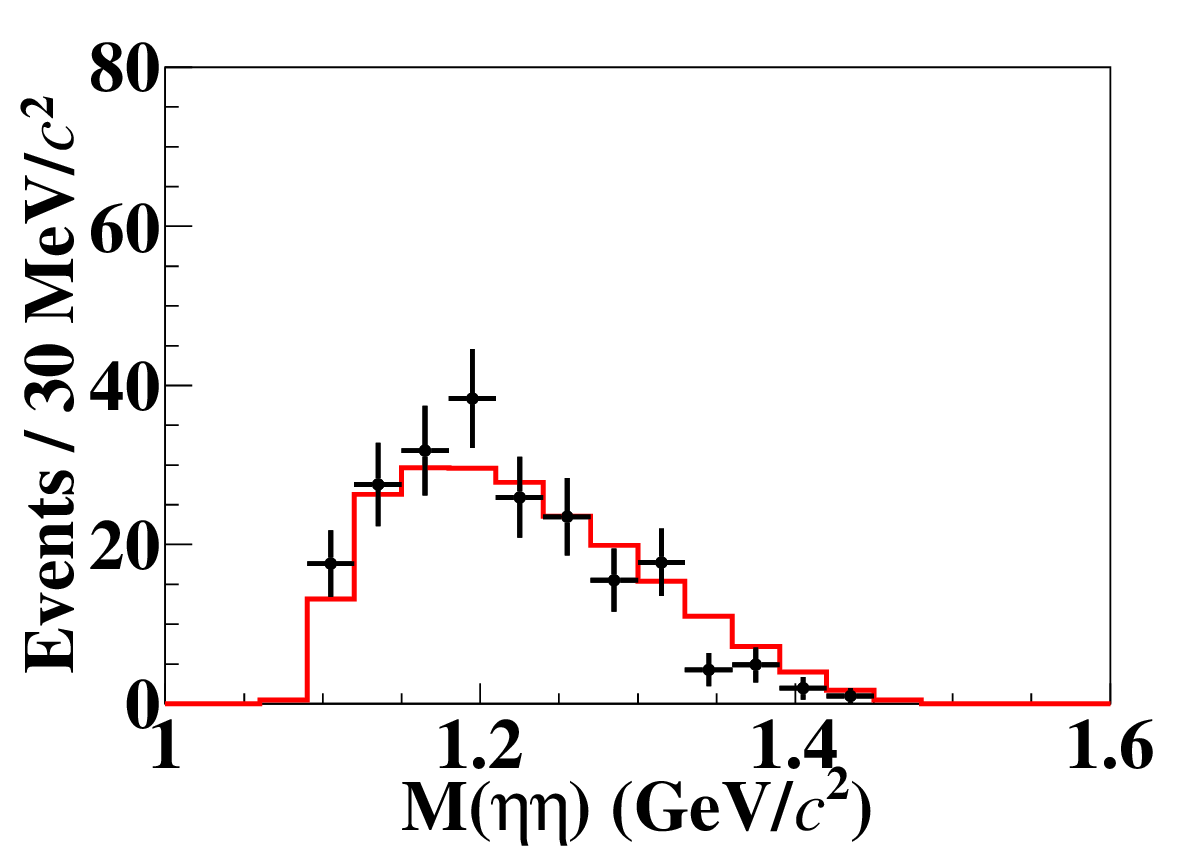}
\end{minipage}
\begin{minipage}[t]{0.32\linewidth}
\includegraphics[width=1\textwidth]{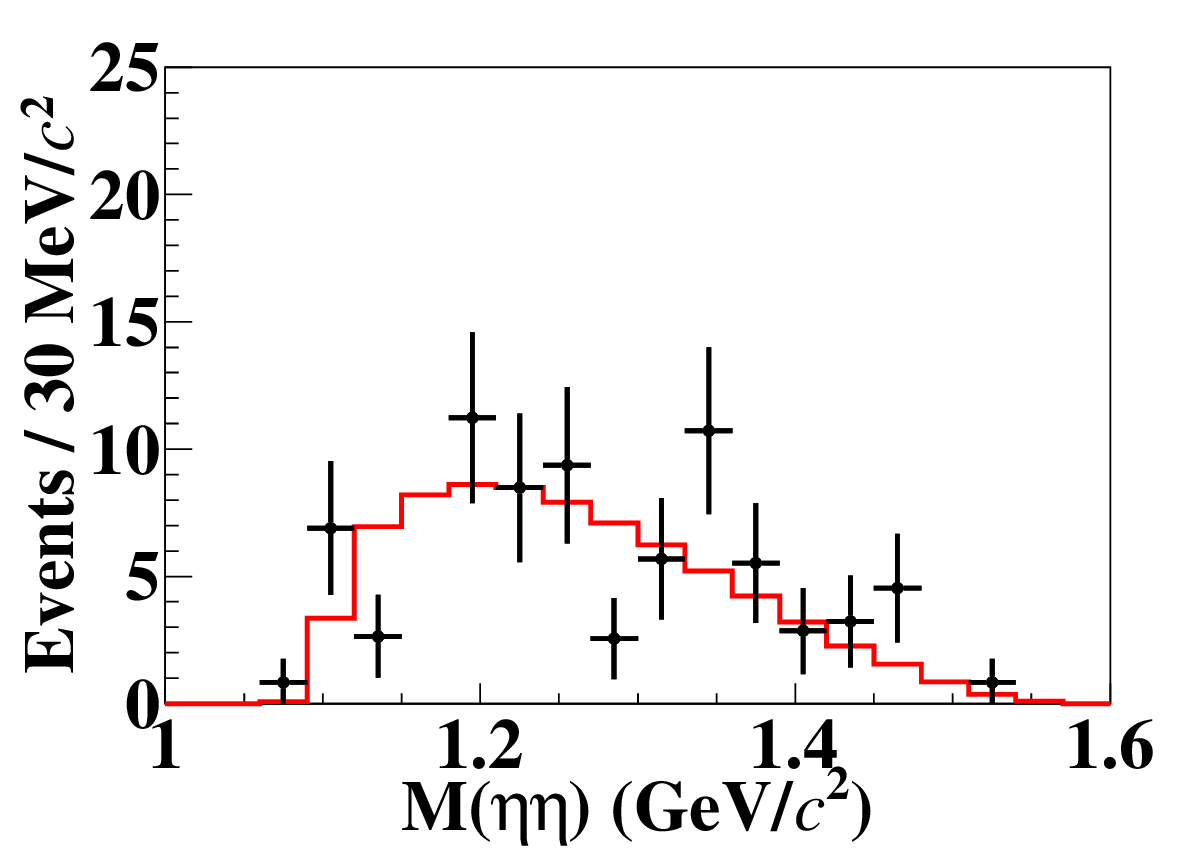}
\end{minipage}
\begin{minipage}[t]{0.32\linewidth}
\includegraphics[width=1\textwidth]{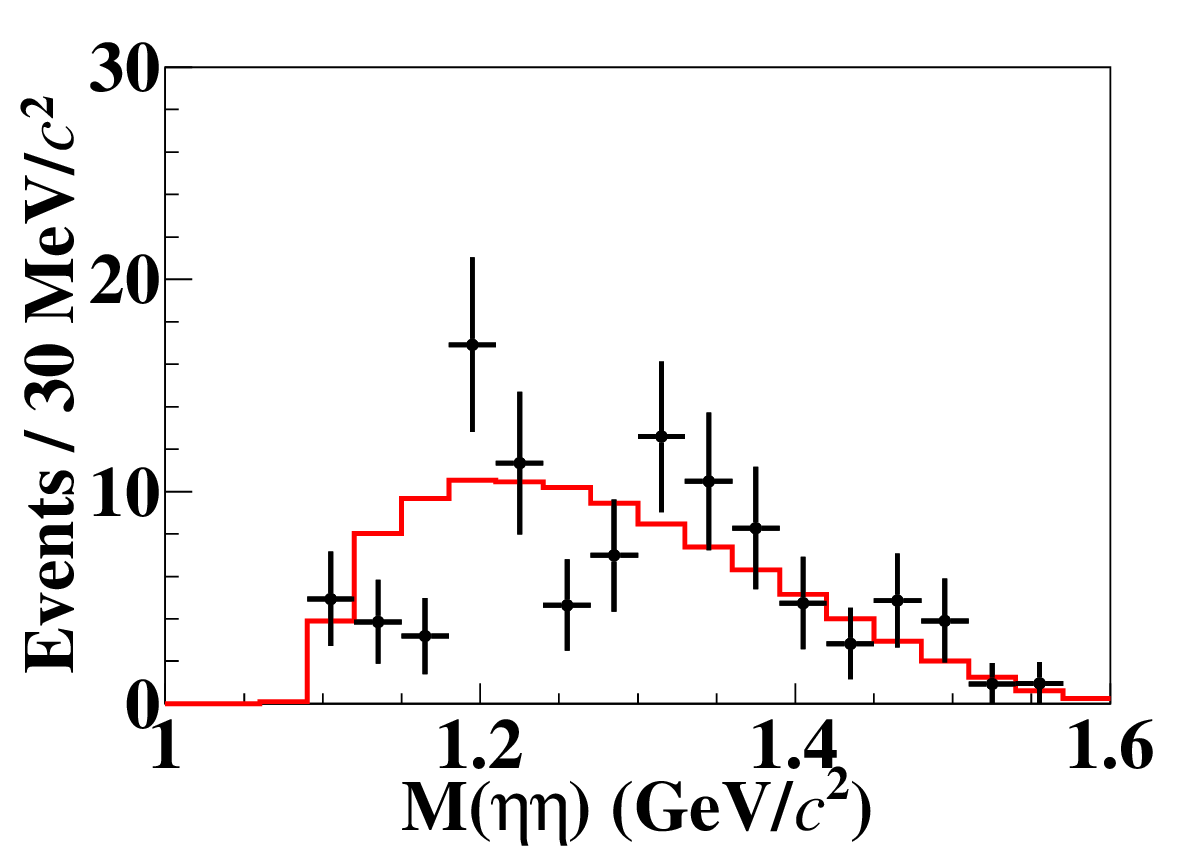}
\end{minipage}

\caption{Invariant mass distributions of $p\eta/\bar{p}\eta$, $\ppb$, and $\eta\eta$ for $\chi_{c0}$ (left),  $\chi_{c1}$ (middle), and $\chi_{c2}$ (right). The points with error bars are data and the histograms are signal MC simulations.}
\label{Compare:BODY3_hists_chic12}
\end{center}
\end{figure*}

Figure~\ref{Compare:BODY3_hists_chic12} shows the distributions for both data and signal MC, no significant structures are observed in all mass spectrums. There are $\chi_{cJ}$ peaking background and smooth background when selecting signal candidates from data, so in these distributions for data, the $\chi_{cJ}$ peaking background is estimated by generating the exclusive MC of the $\chi_{cJ}$ peaking background and normalizing it to the data, and the smooth background is estimated by extracting events from the $\chicJ$ sideband region in $M(\ppb\eta\eta)$ and normalizing them to the signal region. Since there are two $\eta$ candidates, the ones with higher and lower momenta are distinguished by subscripts ``H'' and ``L''. Then the background subtracted $\ppb$, $p\eta/\bar{p}\eta$ and $\eta\eta$ distributions of data are examined for possible intermediate structures, as shown in Fig.~\ref{Compare:BODY3_hists_chic12}.

\section{Numerical results}
The branching fractions of $\chi_{c0,1,2}$ $\to \ppb\eta\eta$ are determined as:
\begin{equation*}
    \Br({\chi_{cJ}} \to p\bar{p}\eta\eta) = \frac{N_{\chi_{cJ}}^{\mathrm{obs}}}{N_{\psi(3686)}\Br(\psi(3686)\too\gamma\chi_{cJ})\Br^{2}(\eta\too\gamma\gamma)\epsilon},
\end{equation*}
where $N_{\chi_{cJ}}^{\mathrm{obs}}$ is the $\chi_{cJ}\to \ppb\eta\eta$ signal yield, $N_{\psi(3686)}$ is the number of $\psi(3686)$ candidates in data~\cite{psip_num_0912}, and $\epsilon$ denotes the detection efficiency obtained from MC simulations. The branching fractions related to the intermediate states are taken from the PDG~\cite{pdg2022}. The measured branching fractions are summarized in Table~\ref{list_summary}.

\section{Systematic Uncertainties}\label{sec:sysU}
The sources of the systematic uncertainty are listed below:

\begin{itemize}

\item[(i)]{\bf Tracking:}
The uncertainty of the tracking for proton or anti-proton is estimated to be 1.0\% for each track by the control sample $J/\psi \too\ppb\pi^+\pi^-$~\cite{Yuan:2015wga}.

\item[(ii)]{\bf Photon reconstruction:}
The uncertainty originating from the photon reconstruction is studied using the control sample $J/\psi \to \rho^0\pi^0$, and is determined to be $1.0\%$ per photon \cite{Ablikim:2010zn}.

\item[(iii)]{\bf PID:}
The uncertainty from the PID for proton or anti-proton is determined to be 1.0\% for each particle by the control sample $J/\psi \to p\bar{p}\pi^+\pi^-$~\cite{Yuan:2015wga}.

\item[(iv)]{\bf Kinematic fit}:
The systematic uncertainty due to the 4C kinematic fit is estimated by comparing the efficiency before and after applying the helix correction \cite{helix}. The correction factors are obtained by studying the control sample $\psipp \to p\bar{p}\pi^0$. The systematic uncertainties are 1.4$\%$, 1.0$\%$ and 1.1$\%$ for $\chi_{c0,1,2}$ channels, respectively.

\item[(v)]{\bf \boldmath$\pi^0$ mass window}:
To evaluate the uncertainty, we perform a Barlow test~\cite{barlow_test} to examine the significant deviation ($\zeta$) between the nominal fit and the systematic test. The variable $\zeta$ is defined as,
\begin{equation}
    \zeta=\frac{\left|V_{\text {nominal }}-V_{\text {test }}\right|}{\sqrt{\left|\sigma_{V\text{nominal }}^2-\sigma_{V\text {test}}^2\right|}},
\end{equation}
where $V$ denotes the measured branching fraction and $\sigma_V^{2}$ is the statistical error of $V$. To obtain the $\zeta$ distribution, we vary the mass window of $\pi^0$ from $\left|M(\gamma\gamma) - m_{\pi^0}\right|>13\text{ MeV}/c^2$ to $\left|M(\gamma\gamma) - m_{\pi^0}\right|>17\text{ MeV}/c^2$ with a step size of $0.5\text{ MeV}/c^2$. Since the values of $\zeta$ are greater than 2, the largest differences relative to the nominal results are taken as the systematic uncertainties, which are 3.1\%, 8.8\% and 7.2\% for $\chi_{c0,1,2}\to \ppb \eta\eta$, respectively.

\item[(vi)]{\bf $\eta$ sideband}:
To estimate the uncertainty from the non-$\eta\eta$ background subtraction via the $\eta$ sideband, we also perform a similar Barlow test~\cite{barlow_test} as above. To obtain the $\zeta$ distribution, we change the interval between the signal and the sideband region from 2.5$\sigma$ to 3.5$\sigma$ with a step size of 0.1$\sigma$. The largest differences relative to the nominal results are taken as the systematic uncertainties, which are 1.2\%, 5.5\%, and 2.5\% for $\chi_{c0,1,2}\to \ppb \eta\eta$, respectively.

\item [(vii)]{\bf Signal shape}:
To estimate the uncertainty caused by the signal shape, we alternatively apply another BW function $(E_{\gamma}^3\times BW(m)\times f_d(E_{\gamma}))\otimes G(\delta m,\sigma)$ to describe the signal shape, where $E_{\gamma}=(m_{\psipp}^2-M_{\ppb\eta\eta}^2)/2m_{\psipp}$~\cite{d3Egamma} is the energy of the transition photon in the $\psipp$ rest frame and $m_{\psipp}$ is the $\psipp$ nominal mass. The function $f_d(E_{\gamma})$ damps the diverging tail raised by $E_{\gamma}^3$, expressed as $E_0^2/(E_{\gamma}E_0+(E_{\gamma}-E_0)^2)$~\cite{fde3}. The difference relative to the nominal BW function is taken as the systematic uncertainty.

\item [(viii)] {\bf Background shape}:
The uncertainty due to the background shape is estimated by replacing the $2^{\rm nd}$-order polynomial function with a $3^{\rm rd}$-order polynomial function. The difference between two background functions is taken as the systematic uncertainty.

\item[(ix)]{\bf Intermediate branching fractions}:
The branching fractions of the intermediate states are obtained from the PDG~\cite{pdg2022}.

\item[(x)]{\bf \boldmath$N_{\psi(3686)}$}:
The uncertainty due to the number of $\psi(3686)$ candidates in data is determined to be 0.5\%~\cite{psip_num_0912}.

\end{itemize}
All the systematic uncertainties on the branching fractions are summarized in Table~\ref{list_sys}. The total systematic uncertainties are obtained by summing each systematic uncertainty in quadrature under the assumption that they are independent.

\begin {table}[htbp]
\setlength{\tabcolsep}{14pt}
\centering
{\caption {Relative systematic uncertainties in the measurements of the branching fractions of $\chi_{cJ}\to \ppb\eta\eta$.}
\label{list_sys}}
\begin {tabular}{l c c c}
\hline
\hline
Source   & $\Delta_{\chi_{c0}}(\%)$ & $\Delta_{\chi_{c1}}(\%)$ & $\Delta_{\chi_{c2}}(\%)$ \\
\hline
Tracking                  & 2.0         & 2.0         & 2.0         \\
Photon reconstruction     & 5.0         & 5.0         & 5.0         \\
PID                       & 2.0         & 2.0         & 2.0         \\
Kinematic fit             & 1.4         & 1.0         & 1.1         \\
$\pi^0$ mass window       & 3.1         & 8.8         & 7.2         \\
$\eta$ sideband           & 1.2         & 5.5         & 2.5         \\
Signal shape              & 0.9         & 0.3         & 0.1         \\
Background shape          & 0.6         & 1.6         & 1.6         \\
Intermediate branching fractions        & 2.4         & 2.8         & 2.5         \\
$N_{\psipp}$              & 0.5         & 0.5         & 0.5         \\
\hline
Total                     & 7.3         &12.3         & 10.1        \\
\hline
\hline
\end{tabular}
\end{table}

\section{Summary}
Based on $(2712.4\pm 14.3)\times 10^6$ $\psi(3686)$ candidates, the decays $\chi_{cJ}(J=0,1,2)\too\ppb\eta\eta$ are observed for the first time with a statistical significances of $13.4\sigma$, $5.4\sigma$, and $9.6\sigma$, respectively. The branching fractions of $\chi_{c0,1,2}\too\ppb\eta\eta$ are determined to be $({5.75\pm0.59\pm0.42}) \times 10^{-5}$, $({1.40\pm0.33\pm0.17}) \times 10^{-5}$, and $({2.64\pm0.40\pm0.27}) \times 10^{-5}$, respectively, where the first uncertainties are statistical and the second systematic. No obvious intermediate states are found in the $p\eta/\bar{p}\eta$, $p\bar{p}$, and $\eta\eta$ systems. In order to further understand the characteristics of $\chicJ$ mesons and excited baryons, the theoretical study of the decay channels of $\chicJ$ remains a key focus of future research.

\acknowledgments
The BESIII Collaboration thanks the staff of BEPCII and the IHEP computing center for their strong support. This work is supported in part by National Key R\&D Program of China under Contracts Nos. 2020YFA0406300, 2020YFA0406400, 2023YFA1606000; National Natural Science Foundation of China (NSFC) under Contracts Nos. 12375071, 11635010, 11735014, 11935015, 11935016, 11935018, 12025502, 12035009, 12035013, 12061131003, 12192260, 12192261, 12192262, 12192263, 12192264, 12192265, 12221005, 12225509, 12235017, 12361141819; Natural Science Foundation of Henan under Contract No. 242300421163; the Chinese Academy of Sciences (CAS) Large-Scale Scientific Facility Program; the CAS Center for Excellence in Particle Physics (CCEPP); Joint Large-Scale Scientific Facility Funds of the NSFC and CAS under Contract No. U1832207; CAS under Contract No. YSBR-101; 100 Talents Program of CAS; The Institute of Nuclear and Particle Physics (INPAC) and Shanghai Key Laboratory for Particle Physics and Cosmology; Agencia Nacional de Investigación y Desarrollo de Chile (ANID), Chile under Contract No. ANID PIA/APOYO AFB230003; German Research Foundation DFG under Contract No. FOR5327; Istituto Nazionale di Fisica Nucleare, Italy; Knut and Alice Wallenberg Foundation under Contracts Nos. 2021.0174, 2021.0299; Ministry of Development of Turkey under Contract No. DPT2006K-120470; National Research Foundation of Korea under Contract No. NRF-2022R1A2C1092335; National Science and Technology fund of Mongolia; National Science Research and Innovation Fund (NSRF) via the Program Management Unit for Human Resources \& Institutional Development, Research and Innovation of Thailand under Contracts Nos. B16F640076, B50G670107; Polish National Science Centre under Contract No. 2019/35/O/ST2/02907; Swedish Research Council under Contract No. 2019.04595; The Swedish Foundation for International Cooperation in Research and Higher Education under Contract No. CH2018-7756; U. S. Department of Energy under Contract No. DE-FG02-05ER41374.

\clearpage

\section*{The BESIII collaboration}
\addcontentsline{toc}{section}{The BESIII collaboration}
\begin{small}
M.~Ablikim$^{1}$, M.~N.~Achasov$^{4,c}$, P.~Adlarson$^{76}$, X.~C.~Ai$^{81}$, R.~Aliberti$^{35}$, A.~Amoroso$^{75A,75C}$, Q.~An$^{72,58,a}$, Y.~Bai$^{57}$, O.~Bakina$^{36}$, Y.~Ban$^{46,h}$
% ...
\\
{\it
$^{1}$ Institute of High Energy Physics, Beijing 100049, People's Republic of China\\
% ...
$^{4}$ Budker Institute of Nuclear Physics SB RAS (BINP), Novosibirsk 630090, Russia\\
% ...
$^{35}$ Johannes Gutenberg University of Mainz, Johann-Joachim-Becher-Weg 45, D-55099 Mainz, Germany\\
$^{36}$ Joint Institute for Nuclear Research, 141980 Dubna, Moscow region, Russia\\
% ...
$^{46}$ Peking University, Beijing 100871, People's Republic of China\\
% ...
$^{57}$ Southeast University, Nanjing 211100, People's Republic of China\\
$^{58}$ State Key Laboratory of Particle Detection and Electronics, Beijing 100049, Hefei 230026, People's Republic of China\\
% ...
$^{72}$ University of Science and Technology of China, Hefei 230026, People's Republic of China\\
% ...
$^{75}$ University of Turin and INFN, (A)University of Turin, I-10125, Turin, Italy; (B)University of Eastern Piedmont, I-15121, Alessandria, Italy; (C)INFN, I-10125, Turin, Italy\\
$^{76}$ Uppsala University, Box 516, SE-75120 Uppsala, Sweden\\
% ...
$^{81}$ Zhengzhou University, Zhengzhou 450001, People's Republic of China\\
\\
$^{a}$ Deceased\\
% ...
$^{c}$ Also at the Novosibirsk State University, Novosibirsk, 630090, Russia\\
% ...
$^{h}$ Also at State Key Laboratory of Nuclear Physics and Technology, Peking University, Beijing 100871, People's Republic of China\\
% ...
}
\end{small}

\end{document}